
\input phyzzx
\def\P{O_I}
\def\Q{O_{\sigma}}
\def\R{O_{\varepsilon}}
\def\L{L^{\phi}_{-1}}
\def\AL{{\overline L^{\phi}_{-1}}}
\def\LL{L^{\phi}_{-2}}

\def\pp{\prime}

\REF\s{N. Seiberg, Prog. Theor. Phys. Supp. {\bf 102} (1990) 319.}
\REF\pa{J. Polchinski, Proc. of the String 1990, Texas A\&M, March 1990.}
\REF\gm{D. Gross and A. Migdal, Phys. Rev. Lett. {\bf 64} (1990) 127;
       Nucl. Phys. {\bf B340} (1990) 333;
       M. Douglas and S. Shenker, Nucl. Phys. {\bf B335} (1990) 635;
       E. Brezin and V. Kazakov, Phys. Lett. {\bf 236} (1990) 144.}
\REF\w{E. Witten, Nucl. Phys. {\bf B340} (1990) 281;
       R. Dijkgraaf and E. Witten, Nucl. Phys. {\bf B342} (1990) 486.}
\REF\fkn{M. Fukuma, H. Kawai and R. Nakayama, Int. J. Mod. Phys. {\bf A6}
       (1991) 1385.}
\REF\dvv{R. Dijkgraaf, E. Verlinde and H. Verlinde, Nucl.Phys. {\bf B348}
       (1991) 435.}
\REF\vv{E. Verlinde and H. Verlinde, Nucl.Phys. {\bf B348} (1991) 457.}
\REF\l{K. Li, preprint CALT-68-1670.}
\REF\pb{J. Polchinski, Nucl. Phys. {\bf B357} (1991) 241.}
\REF\h{K. Hamada, Nucl. Phys. {\bf B365} (1991) 354.}
\REF\dk{J. Distler and H. Kawai, Nucl. Phys. {\bf B321} (1989) 509.}
\REF\bpz{A. Belavin, A. Polyakov and A. Zamolodochikov, Nucl. Phys. {\bf B241}
       (1984) 333.}
\REF\lz{B. Lian and G. Zuckerman, Phys. Lett. {\bf 254B} (1990) 417.}
\REF\ct{T. Curtright and C. Thorn, Phys. Rev. Lett. {\bf 48} (1982) 1309.}
\REF\bk{M. Bershadsky and I. Klebanov, Phys. Rev. Lett. {\bf 65} (1990) 3088.}
\REF\st{N. Sakai and Y. Tanii, Int. J. Mod. Phys. {\bf A6} (1991) 2743.}
\REF\gl{M. Goulian and M. Li, Phys. Rev. Lett. {\bf 66} (1991) 2051.}
\REF\fk{P. DiFrancesco and D. Kutasov, Nucl. Phys. {\bf B342} (1990) 589.}
\REF\cgm{C. Crnkovic, P. Ginsparg and G. Moore, Phys. Lett. {\bf 237B} (1990)
         196.}

\pubnum{91-25}

\titlepage

\title{{\bf Ward Identities of Liouville Gravity coupled to Minimal Conformal
Matter}\footnote\dagger{Talk given at YITP Workshop on Developments in Strings
and Field Theories, Kyoto, Japan, Sept. 9-12 1991.}}

\author{Ken-ji Hamada}

\address{Institute of Physics, University of Tokyo  \break
           Komaba, Meguro-ku, Tokyo, 153, Japan}

\abstract
  The Ward identities of the Liouville gravity coupled to the minimal conformal
 matter are investigated. We introduce the pseudo-null fields and the
generalized equations of motion, which are classified into series of the
Liouville charges. These series have something to do with the W and Virasoro
constraints. The pseudo-null fields have non-trivial contributions at the
boundaries of the moduli space. We explicitly evaluate the several boundary
contributions. Then the structures similar to the W and the Virasoro
constraints
appearing in the topological and the matrix methods are realized. Although
our Ward identities have some different features from the other methods,
the solutions of the identities are consistent to the matrix model results.

\endpage

\chapter{{\bf Introduction}}

  The two dimensional quantum gravity has been studied as a toy model of the
four dimensional quantum gravity. It is important to discuss  what are
common structures independent of the dimension. Recently there are remarcable
developments in this direction\NPrefmark{\s,\pa}. The problem of non linearity
is the one of the most important issues of gravity.
In two dimension the quantum gravity becomes
exactly solvable\NPrefmark{\gm,\w} and the non linear structures such as the
string equations or the W  and the Virasoro
constraints\NPrefmark{\fkn,\dvv,\vv,\l}
are realized in the approaches of matrix and
topological models. To make a comprehension deeper, we reexamine these
structures in terms of the Liouville gravity\NPrefmark{\pb,\h}.
Really the non linear structures  are
directly related to the nature of the Hilbert space\NPrefmark{\s},
the factorization of amplitudes\NPrefmark{\pa,\s} and so on.
Then it is important to discuss whether we have to give
up the superposition principle or not. As we will see later, at least in two
dimension, it appears that there is no need to abandon it.

  In this talk we will discuss how the non linear structures appear in
the Liouville gravity coupled to the minimal conformal field theory (CET). The
quantum Liouville theory has the different features from the standard quantum
field theory. The theory has two kinds of states\NPrefmark{\s}:
microscopic and macroscopic ones.
Microscopic states, which correspond to a branch of the local operators
of Distler-Kawai\NPrefmark{\dk}, are dominated at small area of surface and
are non-normalizable,
while macroscopic states are normalizable and correspond to
macroscopic loops in surface. The existence of two kinds of states is
important when we discuss the non linear structures of the Liouville gravity.

   In Sect.2 we summarize the results of quantum Liouville theory. Here the
macroscopic states are introduced as the Hilbert space of the Liouville theory.
 The intermediate states of amplitudes are expanded by the macroscopic states.
It is natural because these states include informations of fluctuating surface
and are normalizable. The factorization of
matter part is given by BPZ theory\NPrefmark{\bpz} in which the metric on the
space of primary
fields is diagonal. Then we have a question how the non-trivial metric of
scaling operators appearing in the topological and the matrix approaches, or
the structures of W  and Virasoro constraints, are realized from the diagonal
metric of BPZ. This matter is discussed in Sect.4 and 5.

   In Sect.3 we introduce the pseudo-null fields\NPrefmark{\pb,\h},
where \lq\lq pseudo\rq\rq
means that they become exact null fields for the free theory (the cosmological
constant $ \mu = 0 $ ). The pseudo-null fields can be rewritten in
the form of BRST commutator. But it does not mean that they are trivial. We
must  take into account the measure of moduli space. Then the BRST operator
picks
up the non-trivial contributions from various singular boundaries of moduli
space. Thus the pseudo-null fields are essentially non zero and should
satisfy the non-trivial relations, which we could see just as a generalization
of the equation of motion in quantum Liouville theory coupled to the minimal
conformal matter. The pseudo-null fields can be classified in the $ m-1 $
series of the Liouville charge, where the central charge of minimal matter is
$ c_m =1-6/m(m+1) $.

   In Sect.4 and 5 we explicitly evaluate the boundary contributions and derive
 the various Ward identities of two dimensional gravity. Then the factorization
property discussed in Sect.2 and the fusion rule of CFT are
used. The derived equations have similar structures to some of the W  and
Virasoro constraints $L_0 $, $L_1 $ and $W_{-1} $. For the Ising model we can
discuss in detail and derive
a closed set of the Ward identities\NPrefmark{\h}.
The solutions of the identities, which
are summarized in appendix, are consistent to the matrix model results.

   Sect.6 is devoted to conclusions and discussions. We consider the
similarities and the differences between the Liouville gravity and the other
methods. We also discuss another BRST invariant fields found by
Lian-Zuckerman\NPrefmark{\lz}.
Since these fields have the non-standard ghost number, there are some
difficulties when we consider the correlation functions of these fields.

\chapter{{\bf Quantum Liouville Theory}}

   The two dimensional quantum gravity is defined through the functional
integrations over the metric tensor of two dimensional surface $g_{ab} $ and
the matter field $m $. The partition function is
$$
     Z= \sum_{\chi} \kappa^{-\chi} \int [dg_{ab}] [dm]
        \exp \biggl( -{\mu \over 2\pi}
                \int_{M_{\chi}} d^2 z
                 \hbox{$\sqrt{\vert g_{ab} \vert}$} -S_m \biggr)  ~,
     \eqno\eq
$$
where $ S_m $ is a matter action. $\chi $ is the Euler number of surfaces:
$\chi =2-2g $ and
$\kappa $ is the string coupling constant. We use the conformal gauge
$
g_{ab} = {\rm e}^{\gamma \phi} {\hat g}_{ab}({\hat t})
$,
where $\gamma $ is a parameter given below and ${\hat g}_{ab} ({\hat t}) $
is a background metric  parametrized by the moduli ${\hat t} $.
We choose the locally flat
background metric. $\phi $ is well-known as the Liouville field. After fixing
the reparametrization invariance, one can rewrite  the two dimensioal quantum
gravity as
$$
\eqalign{
      <O> & = \sum_g <O>_g                                     \cr
          & = \sum_g  \kappa^{-\chi} \int d^2{\hat t}
                \int [dm d\phi db dc]
              \mu (b) O {\rm e}^{-S_m -S_{\phi} -S_{gh} } ~,   \cr
        } \eqno\eq
$$
where $O $ is some operator. $S_{\phi} $ is the Liouville action
$$
   S_{\phi} = {1 \over 8\pi } \int d^2 z \hbox{$ \sqrt{ {\hat g}}$}
               ({\hat g}^{ab} \partial_a \phi \partial_b \phi
                + Q {\hat R} \phi + 4\mu {\rm e}^{\gamma \phi} )
      \eqno\eq
$$
and $S_{gh} $ is the ghost action.
When the matter system is the minimal conformal field theory with central
charge $c_m =1-6/m(m+1) $, the parameters $Q $ and $\gamma $ are defined as
$$
       Q ={4m+2 \over \sqrt{2m(m+1)}} ~,  \qquad
       \gamma ={2m \over \sqrt{2m(m+1)} } ~.
         \eqno\eq
$$
$\mu (b) $ is the measure of the moduli space. $n $ is the number of
operators.

   The Liouville action has the following scaling property
$$
        S_{\phi} \rightarrow S_{\phi} + {Q \over 2} \chi \delta
            \eqno\eq
$$
when we change the field $\phi $ and the cosmological constant $\mu $ as
$\phi \rightarrow \phi +\delta $, $\mu \rightarrow \mu {\rm e}^{-\gamma\delta}
$. The constant shift of eq.(2.5) can be renormalized into the string coupling
$\kappa$.

\section{{\bf Canonical Quantization of Liouville Theory}}

    To discuss the Hilbert space of quantum Liouville theory we use the
canonical quantization. We first change the variable from the plane coordinate
$z $ to the cylinder one $w=\tau +i\sigma $; $z= {\rm e}^w $.
After Wick rotating $\tau \rightarrow it $,
we reach the Liouville theory in Minkowski space. Then we can  set up the
equal-time commutation relation
$$
     [\phi(\sigma, t), \Pi(\sigma^{\pp}, t)] =i\delta(\sigma -\sigma^{\pp}) ~,
       \eqno\eq
$$
where $\Pi(\sigma, t)={1 \over 4\pi}\partial_t \phi(\sigma,t) $ is the
conjugate momentum. As a normal ordering, we adopt the free field one as used
by
Curtright-Thorn\NPrefmark{\ct}.

    The energy-momentum tensor of the Liouville theory
$ T_{\phi}^{\pm\pm} \equiv {1 \over 2}(T_{\phi}^{00} \pm T_{\phi}^{11} ) $ is
given by
$$   T_{\phi}^{\pm\pm}  = {1 \over 8} ( 4\pi \Pi \pm \phi^{\pp} )^2
                 \mp {Q \over 4} ( 4\pi \Pi \pm \phi^{\pp} )^{\pp}
                   +{\mu \over 2} {\rm e}^{\gamma \phi}
                    +{Q^2 \over 8}    ~.
          \eqno\eq
$$
The Hamiltonian $H_{\phi} =L_{\circ} +{\overline L}_{\circ} $ is
$$
   H_{\phi} = {1 \over 4} \int_0^{2\pi} {d\sigma \over 2\pi}
            \Bigl[ (4\pi \Pi(\sigma) )^2 + \phi^{\pp}(\sigma)^2
            +4\mu {\rm e}^{\gamma \phi(\sigma)} \Bigr] + {Q^2 \over 4} ~.
        \eqno\eq
$$
The prime means the derivative with respect to $\sigma $.
$T_{\phi}^{\pm\pm} $ satisfy the Virasoro algebra only if the parameters $Q $
and $\gamma$ satisfy the relation $ Q={2 \over \gamma} +\gamma $. This relation
 is same as that derived by Distler-Kawai\NPrefmark{\dk}.
The central charge is found to be
$c_{\phi} =1 +3Q^2 $. Also it can be seen that $T_{\phi}^{\pm\pm} $ depends
only on $t \pm \sigma $. The conformal weight of the operator
${\rm e}^{\alpha \phi} $ is shown to be
$h_{\alpha} ={1 \over 2}(\alpha Q - \alpha^2 ) $. Thus in spite of the
interaction the values of $c_{\phi} $ and $h_{\alpha} $ are as if $\phi $ is a
free field. The differece is, as we will see below, that we must take the
branch $\alpha < {Q \over 2}$ out of two solutions of $h_{\alpha} ={1 \over
2}(\alpha Q - \alpha^2 ) $.

\section{{\bf Hilbert Space of Liouville Theory}}

    To manage the interaction term, we simplify discussions by considering the
mini-superspace approximation i.e. ${\rm e}^{\gamma \phi(\sigma) } \rightarrow
 {\rm e}^{\gamma \phi_{\circ} } $, where $\phi_{\circ}= \int_0^{2\pi}d\sigma
\phi(\sigma)/2\pi $ is the mean value of $\phi $. In this approximation, the
Hamiltonian $H_{\phi} =L_{\circ}^{\phi} +{\bar L}_{\circ}^{\phi} $ is simply
$$
    H_{\phi} = -{\partial^2 \over \partial \phi_{\circ}^2 }
               + \mu {\rm e}^{\gamma \phi_{\circ}}
               +{Q^2 \over 4} + N +{\bar N}
     \eqno\eq
$$
where $N $ and ${\bar N} $ are the left and right-moving oscillator levels.
In the case of conformal matter $c \leq 1 $, however, the oscillator modes
are canceled out by the oscillator modes of the ghost and the matter parts.
In fact, when we consider the partition functions on the torus, the Dedekint
$\eta $-functions which come from the determinants of oscillator modes are
canceled out and only the zero mode contributions
survive\NPrefmark{\s,\bk,\st}. The derived results
are exactly same as those of the matrix models. So in the following we do not
attend to the oscillator modes.

   The normalizable wave function for $N={\bar N}=0 $ is given by using the
modified Bessel function as
$$
\eqalign{
    &  H_{\phi}\Psi_p (l) = \biggl( p^2 +{Q^2 \over 4} \biggr)\Psi_p (l)~,  \cr
    &  \Psi_p (l) =\biggl( {2 \over \gamma} p
                        {\rm sinh} {2\pi \over \gamma} p \biggr)^{1 \over 2}
                    K_{2ip/\gamma} (2\hbox{$\sqrt{\mu}$} l/\gamma )         \cr
        }   \eqno\eq
$$
for real $p $, where $l= {\rm e}^{{1 \over 2} \gamma \phi_{\circ}} $.
Since $\Psi_{-p} =\Psi_p $, one can take the region $p >0$. To obtain the wave
function we use the boundary condition
$ \Psi_p \sim {\rm sin}p\phi_{\circ}~(p>0) $  at the limit $l \rightarrow 0 $,
which comes from the fact that the incoming wave completely reflect by the
potential ${\rm e}^{\gamma \phi_{\circ}} $. Note that there is no $p=0 $
ground state. Since the ground state is not included in the Hilbert space, we
cannot define the states by acting the operators on the ground state as in the
standard conformal field theory.

    Now we define the state/operator identification  formally by using
the path integral method just like the Hartle-Hawking wave function
$$
     \Psi_p (l) = \int_D [d\phi] \psi_p (\phi) {\rm e}^{-S} ~,
        \eqno\eq
$$
where $ D $ is the disk with boundary $ \vert z \vert =1$ and the boundary
value of $\phi $ is fixed. The operator $\psi_p (\phi) $ is located
at the centre of the disk $z=0 $, which has the conformal weight
$ h={\bar h}={1 \over 2}p^2 +{Q^2 \over 8} $. Such a operator is given by
$$
     \psi_p (\phi) =\mu^{ip/\gamma} {\rm e}^{(ip+{Q \over 2})\phi}
                    +\mu^{-ip/\gamma} {\rm e}^{(-ip+{Q \over 2})\phi} ~.
          \eqno\eq
$$

   In general the state corresponding to the operator ${\rm e}^{\alpha \phi} $
is  constructed by replacing the operator $\psi_p $ into ${\rm e}^{\alpha \phi}
$ in (2.11). Let us consider the case that $\alpha $ is real, which
corresponds to the operator of Distler-Kawai. In the mini-superspacce
approximation this state behaves like
$ \Psi_{\alpha} ={\rm e}^{(\alpha-{Q \over 2})\phi_{\circ}} $ at
$\phi_{\circ} \rightarrow - \infty $. If we adopt the branch
$\alpha < {Q \over 2} $ as a solution of $h={1 \over 2}(\alpha Q - \alpha^2) $,
the state diverges at $\phi_{\circ} \rightarrow - \infty $. While for the
branch $\alpha >{Q \over 2} $, $\Psi_{\alpha}$ vanishes and gives no
contributions. Therefore we should  take the branch  $\alpha < {Q \over 2} $.
The limit  $\phi_{\circ} \rightarrow - \infty $ corresponds to the small area
 of the surface $g_{ab} = {\rm e}^{\gamma \phi} {\hat g}_{ab} $. So the
state with $\alpha < {Q \over 2} $ is peaked on the small area region. We call
this type of state \lq\lq microscopic state\rq\rq. This state is
non-normalizable. On the other hand, the normalizable eigenstate of the
Liouville Hamiltonian  $\Psi_p $ corresponds to $\alpha =\pm ip+{Q \over 2} $
and oscilates at the small area region. We call it \lq\lq macroscopic
state\rq\rq.

\section{{\bf Factorization of Amplitudes}}

     Let us consider how the correlation function
$ <\prod_i O_{\alpha_i}>_{\Sigma} $  on the Riemann surface $\Sigma $ with
genus $g $ factorizes into the two surfaces $\Sigma_1 $ with genus $g_1 $ and
$\Sigma_2 $ with  $g_2 $. Here $O_{\alpha_i} $ is the operator with the
Liouville charge $\alpha_i $. The total Liouville charges of each part are
$\sum_{i \in \Sigma_1} \alpha_i $ and $\sum_{i \in \Sigma_2} \alpha_i $,
respectively. If they satisfy the normalizability conditions
$$
     \sum_{i \in \Sigma_1 } \alpha_i + {Q \over 2}(2g_1 -1) > 0 ~, \quad
     \sum_{i \in \Sigma_2 } \alpha_i + {Q \over 2}(2g_2 -1) > 0 ~,
       \eqno\eq
$$
the intermediate states are expanded by the normalizable macroscopic states.
We normalize the macroscopic state as
$$
      <{\bar c}c \psi_p \Phi_{\Delta}({\tilde w}=0) ({\bar \partial} {\bar c})
        (\partial c) {\bar c}c \psi_q \Phi_{\Delta^{\pp}}(w=0) >_{g=0}
       = {2 \pi C(p^2) \over \kappa^2 }
             \delta (p-q) \delta_{\Delta,\Delta^{\pp}} ~,
       \eqno\eq
$$
where the two frames $w $ and ${\tilde w} $ are identified as $w{\tilde w} =1
$. $\Phi_{\Delta} $ is the primary field of minimal CFT.
Then the factorization\NPrefmark{\pa,\s} is
$$
\eqalign{
      <O>_{\Sigma} = \sum_{\Delta} \int_{-\infty}^{\infty}
         &   {dp \over 2 \pi}{1 \over  C(p^2)}
            <O_1 ( {\bar \partial}{\bar c} )(\partial c) {\bar c} c
             \psi_p \Phi_{\Delta} (w=0) >_{\Sigma_1}    \cr
        &  \times <{\bar c} c \psi_p \Phi_{\Delta} ({\tilde w}=0)
                  O_2 >_{\Sigma_2}  ~,  \cr
       } \eqno\eq
$$
where we neglect the oscillator modes which do not contribute to the boundary
terms.

\chapter{{\bf Pseudo-Null Fields and Generalized Equations of Motion}}

    Consider the Liouville system as CFT with central charge
$c_{\phi}= 25+6/m(m+1) $. There are several null fields, for example
$$
\eqalign{
       \chi_{1,1}^{\phi} & = L_{-1}^{\phi} \cdot 1 ~,               \cr
       \chi_{1,2}^{\phi}
          & = \biggl( \LL + {m+1 \over m} L_{-1}^{\phi^2}  \biggr)
                 \cdot {\rm e}^{-\sqrt{{m \over 2(m+1)}}\phi}
            \equiv D_{-2}^{(1,2)} \cdot
                     {\rm e}^{-\sqrt{{m \over 2(m+1)}}\phi} ~,      \cr
       \chi_{2,1}^{\phi}
          & = \biggl( \LL + {m \over m+1} L_{-1}^{\phi^2}  \biggr)
                  \cdot {\rm e}^{-\sqrt{{m+1 \over 2m}}\phi}
              \equiv D_{-2}^{(2,1)} \cdot
                     {\rm e}^{-\sqrt{{m+1 \over 2m}}\phi} ~.      \cr
        }   \eqno\eq
$$
In general there exist the null field $\chi_{p,q}^{\phi} $ at the level $pq $
of the primary field ${\rm e}^{\beta_{p,q}\phi} $ with conformal weight
$h_{p,q} $ for $1 \leq p \leq m-1 $, $1 \leq q \leq m $
$$
\eqalign{
      h_{p,q} &=-{1 \over 4m(m+1)} \Bigl\{ \bigl[ p(m+1) +qm \bigr]^2
                         -(2m+1)^2 \Bigr\} ~,                      \cr
      \beta_{p,q} & = {1 \over \sqrt{2m(m+1)}} \bigl[
                          2m+1-p(m+1)-qm \bigr] ~.                \cr
         }  \eqno\eq
$$
We write the null field as
$
\chi_{p,q}^{\phi} = D_{-pq}^{(p,q)} \cdot {\rm e}^{\beta_{p,q} \phi}
$,
where $D_{-pq}^{(p,q)} $ is the proper combination of $L_{-n}^{\phi} ~(n>1) $
with level $pq $. For example see eq.(3.1).

   Now we construct the pseudo-null fields\NPrefmark{\pb,\h}. The first
non-trivial one is
$$
     N_{1,1} = \L \AL \cdot \phi ~.
           \eqno\eq
$$
The dot \lq\lq $\cdot $\rq\rq denotes that the contour surrounds the operator
located on the r.h.s. of it. From the equation of motion $N_{1,1} $ is
proportional to the cosmological constant operator
$$
       N_{1,1} = {\gamma \over 2} \mu {\rm e}^{\gamma \phi} ~.
          \eqno\eq
$$
In the limit $\mu \rightarrow 0 $, the r.h.s. of eq.(3.4) vanishes. So we call
it a pseudo-null field. Note that $N_{1,1} $ field is constructed by modifying
the trivial null field $\chi_{1,1}^{\phi} $ as
$$
       N_{1,1}= {\partial \over \partial \beta}
                 ( \L \AL \cdot {\rm e}^{\beta\phi} ) \vert_{\beta=0} ~.
       \eqno\eq
$$
In the same way we can construct the pseudo-null field corresponding to the
null field $\chi_{p,q}^{\phi} $ as
$$
\eqalign{
      N_{p,q}&= {\partial \over \partial \beta}
                ( D_{-pq}^{(p,q)} {\overline D_{-pq}^{(p,q)}} \cdot
                 {\rm e}^{\beta\phi} \Phi_{p,q} ) \vert_{\beta=\beta_{p,q}} \cr
             &= D_{-pq}^{(p,q)} {\overline D_{-pq}^{(p,q)}} \cdot
                 \phi {\rm e}^{\beta_{p,q}\phi} \Phi_{p,q} )  ~.            \cr
        }   \eqno\eq
$$
Here, to make the physical operator, we combine the Liouville field and the
matter field. $\Phi_{p,q} $ is the primary field of matter system with
conformal dimension
$$
        \Delta_{p,q} ={[p(m+1)-qm]^2 -1 \over 4m(m+1)} ~.
              \eqno\eq
$$
In the analogy of the equation of motion, it is expected that $N_{p,q} $
is proportional to the dressed physical field of $\Phi_{p,q}$
$$
       N_{p,q}= C_{p,q} \mu^{x_{p,q}} {\rm e}^{\alpha_{p,q} \phi} \Phi_{p,q} ~,
          \eqno\eq
$$
where
$$
\eqalign{
     \alpha_{p,q} & ={2m+1 -\vert p(m+1)-qm \vert \over \sqrt{2m(m+1)}} ~, \cr
      x_{p,q} & = {1 \over 2m} [p(m+1)+qm-\vert p(m+1)-qm \vert ] ~.  \cr
         }  \eqno\eq
$$
$C_{p,q} $ is the proportional constant, which have to be determined later.
The exponent of $\mu $ is determined from the scaling property of the Liouville
 action (2.5).

   Since there is the relation $h_{p,q} =h_{m+p,m+1-q} +(m+p)(m+1-q) $, the
null state itself contains a null state. Therefore we can construct the family
of null physical states satisfying the physical state condition
$$
\eqalign{
      1 &= \Delta_{p,q} +h_{p,q} +pq                               \cr
        &= \Delta_{p,q} +h_{m+p,m+1-q} +(m+p)(m+1-q) +pq           \cr
        &= \Delta_{p,q} +h_{2m+p,q} + (2m+p)q +(m+p)(m+1-q) +pq    \cr
        & \cdots                                                   \cr
        }  \eqno\eq
$$
The pseudo-null field  $N_{p,q} $ corresponds to the relation of the first
line. From the second relation we obtain
$$
       M_{p,q} = D_{-pq}^{(p,q)} {\overline D_{-pq}^{(p,q)}}
                  D_{-(m+p)(m+1-q)}^{(m+p,m+1-q)}
                  {\overline D_{-(m+p)(m+1-q)}^{(m+p,m+1-q)}}
                  \cdot \phi {\rm e}^{\beta_{m+p,m+1-q}\phi} \Phi_{p,q} ~,
          \eqno\eq
$$
which is also proportional to the dressed physical field of $\Phi_{p,q} $.
 In the following we write the dressed physical field of $\Phi_{p,q} $ as
$
         O_{p,q} = {\bar c}c {\rm e}^{\alpha_{p,q}\phi} \Phi_{p,q}
$,
where we combine the ghost field.  We also define
$ {\tilde N}_{p,q} = {\bar c}c N_{p,q} $ and
$ {\tilde M}_{p,q} = {\bar c}c M_{p,q} $. Then these fields are summarized
in the table
$$
\matrix{
       n  & = & 0 & 1 & 2  & \ldots & & & &                   \cr
  \beta_n^1 &:& O_1 & - & {\tilde N}_{1,1} & \ldots & {\tilde N}_{1,m}
                &- & {\tilde M}_{1,m} & {\tilde M}_{1,m-1}                 \cr
  \beta_n^2 &:& O_2 & O_{2,1} & -& {\tilde N}_{2,1} & \ldots
         & {\tilde N}_{2,m} &- & {\tilde M}_{2,m}                   \cr
            &  & \vdots & & \ddots & \ddots & \ddots &  & \ddots &     \cr
  \beta_n^{m-1} &:& O_{m-1} &\ldots &\ldots & O_{m-1,1} &-
          & {\tilde N}_{m-1,1} &  \ldots & {\tilde N}_{m-1,m}        \cr
        }  \eqno\eq
$$
Here $O_p = O_{p,p} ~ (p=1, \cdots , m-1) $. The Liouville charges
$\beta_n^p ~ (p=1, \cdots, m-1) $ are given by
$$
     \beta_n^p = -{(n+p-3)m+p-1 \over \sqrt{2m(m+1)}}  \quad
                   (n \not= p+1 \hbox{ mod } m+1)
           \eqno\eq
$$
It is expected that the series $\beta_n^1 $ have something to do with the
Virasoro constraints and $\beta_n^i ~(i=2, \cdots,m-1) $ with W constraints.
In the following section we dicuss this correspondence.

   The pseudo-null field can be rewritten in the form of BRST commutator
$
   {\tilde  N}_{p,q}=Q^B \cdot W_{p,q}
$, for example $ W_{1,1} = Q_s b_{-1} {\bar b_{-1}} \cdot {\bar c} c \phi $,
where $Q^B =Q_s +{\bar Q}_s $ is the BRST charge.
Thus the generalized equation of motion (3.8) can be rewritten as
$$
         C_{p,q} \mu^{x_{p,q}} O_{p,q} =Q^B \cdot W_{p,q} ~,
              \eqno\eq
$$
In the following section we consider the Ward identity which is given by
inserting the identity (3.14) into the correlator of scaling operators.

\chapter{{\bf Ward Identities of 2D Quantum Gravity  \break
                                coupled to the Ising Model}}

   Let us first discuss the case of the Ising model\NPrefmark{\h}.
We derive various Ward
idetities obtained by inserting the pseudo-null field relations (3.14) into
the correlation functions:
$
<O>_g = < \prod^{n_1} \P \prod^{n_2} \Q \prod^{n_3} \R >_g
$,
where $ \P =O_{1,1} $ is the cosmological constant operator. $ \Q=O_{1,2} $
and $\R=O_{2,1} $ are the dressed spin and energy operators.

\section{{\bf Equation of Motion}}

    The pseudo-null field relation of $N_{1,1} $ is nothing but the equation
of motion. In this subsection we treat this operator. As evaluating the
boundary contributions, there is a problem. The operator $W_{p,q} $ of
eq.(3.14) is in general not well-defined on the moduli space, or it is not
annihilated by the action of $b_{\circ} $ and ${\bar b}_{\circ} $.
Therefore we introduce the well-defined operator $X_{p,q} $ satisfying the
condition
$
b_{\circ} \cdot X_{p,q} ={\bar b}_{\circ} \cdot X_{p,q} = 0
$.
The operator $X_{1,1} $ is defined by modifying $W_{1,1} $ slightly as
$
X_{1,1} =\L {\bar b}_{-1} \cdot {\bar c}c \phi
$.
Then the identity (3.14) for the Ising case becomes
$$
        {3 \over 2\sqrt 6 } \mu \P = Q^B \cdot X_I + K_I  ~.
            \eqno\eq
$$
where $X_I =X_{1,1} $ and $K_I $ is called \lq\lq ghost pieces\rq\rq:
$
K_I = -{7 \over \sqrt 6 } c_{-1} {\bar b}_{-1} \cdot {\bar c} c
$,
which has the non-standard asymmetric ghost number.

   The Ward identity we discuss is
$$
      {3 \over 2\sqrt 6} \mu < \P O>_g  =  < Q^B \cdot X_I O >_g
          + < K_I O >_g ~.
         \eqno\eq
$$
The correlation function with the ghost piece $K_I $ vanishes
because it has the
non-standard ghost number. The first term on r.h.s.  is evaluated as
follows. Taking into account the moduli and the measure for the position
$z=z_1 $ of $ Q^B \cdot X_I $, we obtain
$$
     \int d^2 z_1 b_{-1} {\bar b}_{-1} Q^B \cdot X_I (z_1) O_{\alpha}(z_i)=
    -{1 \over 2 i }  \oint_{\vert z_1 -z_i \vert = \epsilon } d z_1
        \partial \phi (z_1) O_{\alpha}(z_i) ~.
       \eqno\eq
$$
Here the BRST algebra $ \{ Q^B , b_{-1} \} = L_{-1} =\partial $ is used. The
r.h.s. of eq.(4.3) becomes a total derivative with respect to the moduli
so that finite contributions will come from the boundary of moduli space.
In this case the relevant boundary is where $Q^B \cdot X_I (z_1) $ approaches
other operators $O_{\alpha} (z_i) $. To evaluate the boundary term the small
cut-off $\epsilon $ is introduced and, after the calculation, we take the
limit $\epsilon \rightarrow 0 $. The contributions come from
the singularity of operator product $\partial \phi (z)
O_{\alpha}(z_i) $. The leading singularity comes from the free field
$(\mu =0) $ OPE. As the next leading singularity a $\mu $-dependent term
appear,
but, in this case, does not contribute because the power of singularity is too
small to give the finite value at the limit $\epsilon \rightarrow 0 $.

  We also have to analyze curvatures carefully.
One can choose a metric which is almost flat except for delta function
singularities at the positions of scaling operators; $\sqrt g R =4\pi \sum_i
\nu_i \delta^2 (z-z_i) $, so that $ \sum_i \nu_i =\chi $,
where $ \chi $ is the Euler number of two dimensional surface with genus $g $:
$ \chi =2-2g $. We do not assign the curvature to $Q^B \cdot X_I $. To use
free field operator products it is necessary to smooth out the curvature
singularity in the neighborhood of the position of the operator. This is done
by the coordinate transformation:
$
z- z_i =(z^{\pp}-z^{\pp}_i )^{1-\nu_i}
$, or
$
dz d{\bar z} \sim {dz^{\pp} d{\bar z}^{\pp} \over
          \vert z^{\pp} - z^{\pp}_i \vert^{2\nu_i} }
$.
In the smooth $z^{\pp} $-frame we can freely use the operator products.

    After evaluating OPE, we finally obtain the expression
$$
\eqalign{
   \mu {\partial \over \partial \mu }
    & < \prod^{n_1} \P \prod^{n_2} \Q \prod^{n_3} \R >_g  =
    - {\mu \over 2\pi} < \prod^{n_1+1} \P \prod^{n_2} \Q \prod^{n_3} \R >_g \cr
   & = -\biggl( n_1 +{5 \over 6}n_2 +{1 \over 3}n_3 - {7 \over 6} \chi \biggr)
        < \prod^{n_1} \P \prod^{n_2} \Q \prod^{n_3} \R >_g ~.    \cr
        } \eqno\eq
$$
Thus the $\mu $-dependence of the correlation
functions\footnote\dagger{$Z^{g=0}_{n_1,n_2,n_3} $ can be directly calculated
by using CFT methods\NPrefmark{\gl}} is
$$
       < \prod^{n_1} \P \prod^{n_2} \Q \prod^{n_3} \R >_g  =
         Z^g_{n_1, n_2, n_3}
          \mu^{{7 \over 6} \chi -n_1 -{5 \over 6}n_2 -{1 \over 3}n_3 } .
     \eqno\eq
$$
Note that, since the path integral of Liouville field diverges, the derivation
can not apply for the case
$\sum^{n_1 +n_2 +n_3 }_{i=1} \alpha_i -{Q \over 2} \chi < 0 $, where
$\alpha_i $ is the charge of the exponential operator.  Therefore in this case
we have to define the correlation function by using the differential equation
such as, for example,
$ -2\pi {d \over d\mu} <\Q\Q>_{\circ} = <\Q\Q\P>_{\circ} $.

\section{{\bf Ward Identities corresponding to Virasoro Constraints}}

  In this section we consider the Ward identities given by inserting the
pseudo-null field $N_{1,2} =N_{\sigma} $. In the following we consider the
Ward identity
$$
    C_a \mu^{4/3} <\Q \prod_{\alpha} O_{\alpha} >_g =
     <Q^B \cdot X_{\sigma} \prod_{\alpha} O_{\alpha} >_g  +
        \hbox{ghost term,}
    \eqno\eq
$$
where $O_{\alpha} $'s are only the \lq\lq gravitational" primary
fields $\P $ and $\Q $. $X_{\sigma} $ is the well-defined operator on moduli
space:
$
X_{\sigma} = D_{-2}^{\sigma} {\overline B}_{-2}^{\sigma}
      \cdot {\bar c}c \phi {\rm e}^{-{3 \over 2\sqrt 6}\phi} \sigma
$.
To evaluate the r.h.s. of eq.(4.6) we must take into acount the 3 types of
boundaries.

  As discussed in Sect.4.1 the first boundary arises from that $X_{\sigma} $
approaches other operators $O_{\alpha} $.  The relevant operator product is
$$
\eqalign{
        \int_{\vert z_1 - z_i \vert \geq \epsilon}   &
         d^2 z_1 b_{-1} {\bar b}_{-1} Q^B \cdot X_{\sigma} (z_1)
           O_{\alpha} (z_i)      \cr
        & = -{1 \over 2 i} \oint_{\vert z_1 - z_i \vert = \epsilon}
          (dz_1 b_{-1} + d{\bar z}_1 {\bar b}_{-1} ) \cdot
           X_{\sigma} (z_1) O_{\alpha} (z_i) ~.    \cr
        } \eqno\eq
$$
One can see that from the power of OPE singularities between
$ {\bar b}_{-1} \cdot
X_{\sigma} $ and $O_{\alpha} $ the integral of $d {\bar z} $ vanishes. The
integral of $d z $ gives the finite contributions for $\alpha = \sigma $.
Then we get
$$
     -{1 \over 2i} \oint dz b_{-1} \cdot
               X_{\sigma} (z_1) O_{\sigma}(z_i) =
                {5 \pi \over 9\sqrt{6}} O_{\varepsilon}(z_i) ~.
           \eqno\eq
$$

  In the interacting theory there will be the next leading
OPE singularities which
depend on the cosmological constant $\mu $. If one uses the  free field OPE,
one should add the $\mu $-dependent term, or the operators $O_{\alpha} (\alpha
= I, \sigma )$ behave in the neighborhood of boundaries as follows
$$
\eqalign{
   &  O_I \rightarrow  {\bar c} c ( {\rm e}^{{3 \over \sqrt 6}\phi}
             + \eta_I \mu^{1/3} {\rm e}^{{4 \over \sqrt 6}\phi} ) ~,       \cr
   &  O_{\sigma} \rightarrow  {\bar c} c ( {\rm e}^{{5 \over 2\sqrt 6}\phi}
             + \eta_{\sigma} \mu^{2/3} {\rm e}^{{9 \over 2\sqrt 6}\phi} ) ~.
        \cr
         } \eqno\eq
$$
The charge of the Liouville mode and the power of $\mu $ are determined from
the ristriction of conformal dimension and the scaling symmetry of the
Liouville action. The phases $\eta_I $ and $\eta_{\sigma} $ are determined
by the consistency.

   Next we discuss the boundary-2 where the field $X_{\sigma} $ approaches
the pinched point which divides the surface into two pieces. Then the
factorization discussed in Sect.2 is important.
$$
\eqalign{
      <Q^B \cdot X_{\sigma} & O>_{\Sigma}
           = \sum_{\Delta}  \int^{\infty}_{-\infty}
            {dp \over 2\pi}{1 \over C(p^2)}
            <O_1 \int {d^2 z_1 \over z_1 {\bar z}_1 }
           b_{\circ} {\bar b}_{\circ}    Q^B \cdot X_{\sigma}(z_1)
           \int_{\vert q \vert \geq \epsilon } {d^2 q \over q {\bar q} }
           b_{\circ} {\bar b}_{\circ}                       \cr
         &  \times  q^{L_{\circ}} {\bar q}^{{\bar L}_{\circ}}
             ( {\bar \partial}{\bar c} )(\partial c) {\bar c} c
             \psi_p \Phi_{\Delta} (w=0) >_{\Sigma_1}
              <{\bar c} c \psi_{-p} \Phi_{\Delta} ({\tilde w}=0)
                  O_2 >_{\Sigma_2}  ~,  \cr
       } \eqno\eq
$$
where $\Phi_{\Delta}~ =~ I, ~\sigma,~ \varepsilon $.
The coordinates $w $ and ${\tilde w} $ are
defined in the neighborhood of the nodes of $\Sigma_1 $ and $\Sigma_2 $. These
are identified as $w {\tilde w} =q $, where $w=z $ and $q $ is the moduli that
determines the shape of the pinch. By this identification the normalization
(2.14) changes so that the operator $q^{L_{\circ}} {\bar q}^{{\bar L}_{\circ}}
$ is inserted. We explicitly introduce the measure of moduli for $z_1 $ and
$q $. The operator $b_{\circ} $ is defined by the contour integral
around $w=0 $.
Using the BRST algebra one can rewrite the expression into the derivatives
with respect to the moduli $q $.

  The boundary contributions come from the limit that $z_1 $ and $ q $ approach
 zero simultaneously. In this limit the integrand is highly peaked and we can
evaluate the
integral by the saddle point method. The final result becomes
$$
\eqalign{
       <Q^B \cdot X_{1,2} \prod_{j \in S} O_j >_g^{b\sharp 2}
        \simeq  \sum_{S={\overline X} \cup {\overline Y} \atop g=g_1 +g_2}
           \Bigl\{ &
          <O_{\sigma} \prod_{\alpha \in {\overline X}} O_{\alpha} >_{g_1}
             <O_I \prod_{\alpha \in {\overline Y}} O_j >_{g_2}        \cr
         & + <O_I \prod_{\alpha \in {\overline X}} O_{\alpha} >_{g_1}
             <O_{\sigma} \prod_{\alpha \in {\overline Y}} O_j >_{g_2}
                   \Bigr\}  ~, \cr
        }   \eqno\eq
$$
where $S={\overline X} \cup {\overline Y} $ means that the sum is over the
posible factorizations satisfying the conditions (2.13). To derive these
structures we use the fusion rule of the minimal CFT. Here we neglect the
curvature contributions. Note that the metric we
first introduceed (4.10) is the diagonal one, but after evaluating the boundary
the
metric structure changes to the asymmetric form and $\R $ disappears such  as
the topological and the matrix models. This structure really corresponds to the
 $L_1 $ Virasoro constraints.

   The boundary-3 is a kind of boundary-2, where a
handle is pinched. In this case the surface is not divided by the pinching.
Thus we obtain
$$
         <Q^B \cdot X_{1,2} \prod_{j \in S} O_j >_g^{b\sharp 3}
           \simeq  <\P \Q \prod_{\alpha} O_{\alpha}>_{g-1} ~.
              \eqno\eq
$$

   Let us consider the Ward identity with operator insertion $\prod O_{\alpha}
= \Q \prod^n \P $.
In general genus we get $(n \geq 3 $ for $g=0 )$
$$
\eqalign{
    0 = &{5 \over 9\sqrt 6}  \pi <\R \prod^n \P >_g      \cr
       & - 2 \lambda \sum^g_{g_1=0} \sum^n_{k=0}
          \biggl\{
           {n-2 \choose k} \Bigl( {5 \over 12} \chi_1 -{1 \over 3} \Bigr)^2
           +{n-2 \choose k-2} \Bigl( {5 \over 12} \chi_2 -{1 \over 2} \Bigr)^2
              \cr
      & \qquad \quad
             -2{n-2 \choose k-1} \Bigl( {5 \over 12} \chi_1 -{1 \over 3} \Bigr)
               \Bigl( {5 \over 12} \chi_2 -{1 \over 2} \Bigr)
           + {25 \over 144} {n-2 \choose k-1} \sum_i \nu^2_i
          \biggr\}       \cr
      & \qquad \qquad \qquad \quad
           \times <\Q \Q \prod^k \P >_{g_1} <\prod^{n-k+1} \P >_{g_2}  \cr
      & -{\lambda \over 72} <\Q \Q \prod^{n+1} \P>_{g-1}
            + \hbox{ghost term.}      \cr
        }  \eqno\eq
$$
Here the curvature singularities are considered, which are assigned to $n $
$\P $ operators\footnote\dagger{Although the expression changes by how to
assign the curvatures, the final results are independent of the assignments.
Furthermore
as a consistency check we can see that the expression (4.13) is indeed
independent of the value of $\sum_i \nu_i^2 $ for $n \geq 3 $, $g=0 $.}.
We determine the unknown constants except
$\lambda ={8\pi^2 \over C(p^2=-1/6)} $ from the consistency of the Ward
identities on the sphere. In the end the next leading terms of the boundary-1
and $C_a $-term are absorbed in the factorization form.
In the above expression it appears as if
there were no restrictions like the inequalities (2.13).

    In general genus  we need the contributions of the ghost term. Naively
it does not contribute because of the asymmetry of the ghost number. However,
for $g \geq 1$, there will be the non-zero contribution when we evaluate the
curvature singularities. To calculate the curvature contributions we used
the transformation from the
singular frame to the non-singular frame. Then the mapping analytic in the
moduli was used. This is correct on the sphere,
but for $g \geq 1$ one can not take  such a mapping globally to remove the
curvature singularities.
So  there are posibilities that the measure makes up for the asymmetry
and the ghost term contributes.  Really eq.(4.13)
is inconsistent if there are no contributions of the ghost term.
Exceptional case is $g = 1$, then we can choose the flat metric where all
$\nu_i $'s are zero. In this case the ghost term will vanish.

   We also cosider the Ward identity with $n $ $\Q $ operators. The curvature
singularities are assigned to $n $ $\Q $ operators. Then we obtain
$$
\eqalign{
     0 = & {5 \over 9\sqrt 6}  \pi
       \Bigl( n- \chi -{5 \over 4} \sum_i \nu^2_i \Bigr)
           <\R \prod^{n-1} \Q >_g      \cr
        & -  2 \lambda \sum^g_{g_1=0} \sum^n_{k=0}
          \biggl\{
           {n-2 \choose k} \Bigl( {5 \over 12} \chi_1 -{1 \over 3} \Bigr)^2
           +{n-2 \choose k-2} \Bigl( {5 \over 12} \chi_2 -{1 \over 2} \Bigr)^2
              \cr
      & \qquad \quad
             -2{n-2 \choose k-1} \Bigl( {5 \over 12} \chi_1 -{1 \over 3} \Bigr)
               \Bigl( {5 \over 12} \chi_2 -{1 \over 2} \Bigr)
           + {25 \over 144} {n-2 \choose k-1} \sum_i \nu^2_i
          \biggr\}       \cr
      & \qquad \qquad \qquad \quad
           \times < \prod^{k+1} \Q >_{g_1} < \P \prod^{n-k} \Q >_{g_2}  \cr
      & -{\lambda \over 72} <\P \prod^{n+1} \Q>_{g-1}
            + \hbox{ghost term.}      \cr
        }  \eqno\eq
$$

\section{{\bf Ward Identities corresponding to W  constraints}}

  In this section we consider the case of the pseudo-null field
$N_{2,1} =N_{\varepsilon}$:
$$
    C_1 \mu <\R \prod_{\alpha} O_{\alpha} >_g =
     <Q^B \cdot X_{\varepsilon} \prod_{\alpha} O_{\alpha} >_g  +
        \hbox{ghost term,}
    \eqno\eq
$$
where $\alpha ~=~ I, ~ \sigma $. The operator $X_{\varepsilon}=X_{2,1} $ is
defined as in the previous section by
$
X_{\varepsilon}  = D^{\varepsilon}_{-2} {\overline B}^{\varepsilon}_{-2}
                     \cdot {\bar c} c \phi
                  {\rm e}^{-{2 \over \sqrt 6} \phi } \varepsilon
$.
In this case we have to take into account the 4 types of boundaries.

   We do not repeat the calculations in detail. The second and third boundary
contributions have the following form
$$
   b\sharp2,3 \simeq
          \sum_{S={\overline X} \cup {\overline Y} \atop g=g_1 +g_2}
          <\Q \prod_{\alpha \in {\overline X}} O_{\alpha} >_{g_1}
             <\Q \prod_{\alpha \in {\overline Y}} O_j >_{g_2}
           + <\Q \Q \prod_{\alpha \in S} O_{\alpha} >_{g-1} ~.
      \eqno\eq
$$
The operator $\R $ does not appear on the nodes. The metric structure
really corresponds to the $W_{-1} $ constraint.

    Furthermore we must take into account the boundary-4 that
$X_{\varepsilon} $ and two $\Q $'s approach  at a point
simultaneously. In fact one can easily see that, if there is no boundary
contribution of this type, the Ward identity becomes inconsistent. We do not
know how to evaluate this boundary directly.
Instead we assume the following form
$$
        Q^B \cdot X_{\varepsilon} \Q \Q \rightarrow  \P  ~.
        \eqno\eq
$$

    Let us consider the Ward identity of the type:
$
\prod_{\alpha} O_{\alpha}=  \Q \Q \prod^n \P
$.
If the curvatures are assigned only to $n $ cosmological constant operators, we
 obtain the following Ward identity
$$
\eqalign{
        0 = &{2 \pi \over \sqrt 6}   \Bigl(
             \mu {\partial \over \partial \mu} +n +{11 \over 18} \chi
               -{55 \over 18} \sum_i \nu^2_i  \Bigr)
                  < \R \Q \Q \prod^{n-1} \P >_g                \cr
            & +{1 \over 36} \lambda
                \sum^g_{g_1 =0} \sum^{n}_{k=0} \biggl\{ {n \choose k}
                   -55 \Bigl[ {n-2 \choose k} (1-\chi_1)^2        \cr
            & \quad
                -2 {n-2 \choose k-1} (1-\chi_1)(1-\chi_2)
                + {n-2 \choose k-2} (1-\chi_2)^2
                + {n-2 \choose k-1} \sum_i \nu^2_i \Bigr] \biggr\} \cr
            & \qquad \qquad \quad \times
                  < \Q \Q \prod^k \P >_{g_1}
                   < \Q \Q \prod^{n-k} \P >_{g_2}                \cr
            & + {1 \over 72} \lambda
                     <\prod^4 \Q \prod^n \P >_{g-1}
              + {32 \over 3} \pi^2 C_4  < \prod^{n+1} \P >_g
              + \hbox{ghost terms}                             \cr
         }  \eqno\eq
$$
For the case with the operator insertions
$\prod_{\alpha} O_{\alpha} = \prod^n \Q $, if the curvatures are assigned
only to $n $ $\Q $, we get
$$
\eqalign{
        0 = & {1 \over \sqrt 6}  \mu < \R \prod^n \Q >_g       \cr
            & -{1 \over 72}\lambda   \sum^g_{g_1=0}
                \sum^n_{k=0} \biggl\{ {n \choose k}
                   -55 \Bigl[ {n-2 \choose k} (1-\chi_1)^2         \cr
            & \quad
                -2 {n-2 \choose k-1} (1-\chi_1)(1-\chi_2)
                + {n-2 \choose k-2} (1-\chi_2)^2
                + {n-2 \choose k-1} \sum_i \nu^2_i \Bigr] \biggr\} \cr
            & \qquad \qquad \quad \times
                  < \prod^{k+1} \Q >_{g_1}
                   < \prod^{n-k+1} \Q >_{g_2}                \cr
            & - {1 \over 72} \lambda
                     < \prod^{n+2} \Q >_{g-1}
              - {32 \over 3} \pi^2 C_4 \biggl\{ {n(n-1) \over 2}
                     -{77 \over 60}(n-1) \chi +{11 \over 24} \chi^2   \cr
            & \qquad \qquad
                  +{11 \over 48}(n-3) \sum_i \nu^2_i  \biggr\}
                        <\P \prod^{n-2} \Q >_g
                   + \hbox{ghost terms.}                             \cr
         }  \eqno\eq
$$

   Unfortunately we cannot determine the constants $\lambda $ and $C_4 $ by
the consistency. The determination of these values and the ghost terms
remaines  as future problems. We could determine these values by using
the results of ref.18, which are given by $\lambda ={2 \over \sqrt{6}} \pi $
and $ C_4 ={5 \over 6\sqrt{6}}{1 \over \pi} $. Then we can derive several
correlation functions on the sphere and the torus, which are consistent to the
results of the two matrix model\NPrefmark{\cgm}.

\chapter{{\bf Ward Identities for Minimal CFT}}

    In the previous section we obtain a closed set of Ward identities for
the case of the Ising model. For the general minimal series it is difficult to
derive a closed set of Ward identities because  more complicated boundaries
contribute and also the number of primary fields increases. So we only
concentrate on the pseudo-null field $N_{1,2} $. Then it is expected that the
structures like $L_1 $ equation, or metric on the space of scaling operators
appearing in the matrix and the topological methods, are realized.

   The Ward identity is given by substituting the relation (3.14) with
$(p,q) = (1,2) $ into correlation functions. For simplicity we consider the
correlation function with the operators $O_j \equiv O_{j,j}~(j=1,\cdots ,m-1)$,
 which corresponds to the gravitational primary fields. Then
$$
    C_{1,2} \mu^{{m+1 \over m}} <O_{1,2} \prod_{j \in S} O_j >_g
      = <Q^B \cdot W_{1,2} \prod_{j \in S} O_j >_g ~.
       \eqno\eq
$$
The operator $O_{1,2}=O_{m-1,m-1}$ corresponds to the first gravitational
primary $O_{m-1} $.
We evaluate the boundary contributions of the r.h.s. of
eq.(5.1). The contributions of the first boundaries are given by
$$
    <Q^B \cdot W_{1,2} \prod_{j \in S} O_j >_g^{b\sharp 1}
        \simeq \sum_{k \in S \atop (k \not= 1)}
           <O_{k,k-1} \prod_{j (\not=k)} O_j >_g
           + \hbox{ next leading terms,}
       \eqno\eq
$$
where $O_{k,k-1} $ corresponds to the gravitational descendant
$\sigma_1 (O_k ) ~(k=2,\cdots, m-1) $. Here we neglect the curvature
contributions and the normalization of scaling operators. The contributions
from the boundaries-2 and -3 are given by
$$
\eqalign{
     <Q^B \cdot W_{1,2} \prod_{j \in S} O_j >_g^{b\sharp 2,3}
        & \simeq \sum_{k=1}^{m-1} \biggl\{
               <O_k O_{m-k} \prod_{j \in S} O_j >_{g-1}       \cr
       & + \sum_{S={\overline X} \cup {\overline Y} \atop g=g_1 +g_2}
          <O_k \prod_{j \in {\overline X}} O_j >_{g_1}
             <O_{m-k} \prod_{j \in {\overline Y}} O_j >_{g_2}   \biggr\} ~, \cr
         }  \eqno\eq
$$
where $S={\overline X} \cup {\overline Y} $ means that the sum is over the
posible
factorizations satisfying the conditions (2.13). To derive these structures we
use the fusion rule of the minimal CFT. As discussed in the case of the Ising
model, the l.h.s. of eq.(5.1) and the next leading terms of eq.(5.2) are used
to complete the factorization form of eq.(5.3). Then we finally obtain the
following structure
$$
\eqalign{
   0 \simeq & \sum_{k \in S \atop (k \not= 1)}
           <O_{k,k-1} \prod_{j (\not=k)} O_j >_g               \cr
            & + \sum_{k=1}^{m-1} \biggl\{
                 <O_k O_{m-k} \prod_{j \in S} O_j >_{g-1}       \cr
            & \quad + \sum_{S= X \cup Y \atop g=g_1 +g_2}
              <O_k \prod_{j \in  X} Oj >_{g_1}
                  <O_{m-k} \prod_{j \in Y} O_j >_{g_2}   \biggr\} ~, \cr
         }  \eqno\eq
$$
The equation really has the similar structures to the $L_1 $ Virasoro
constraint. The metric on the space of scaling operators appearing in
the other methods are realized explicitly. It is essentially determined by
the fusion rule of CFT and the conservation of the total Liouville charge.

   It probably needs to discuss all $N_{p,q} $ fields to determine all
correlation functions of $O_{p,q} $. It appears, however, that $M_{p,q} $ and
the others do not
give the essentially new informations as far as one considers only the
correlation functions of the operators in the Kac table $O_{p,q} $.

    The difference from the other methods is that the l.h.s. of eq.(5.4)
vanishes and the identity
is closed only by the operators $O_{p,q} $ in the Kac table, while $L_1 $
equation appearing in the other methods has the structure that the l.h.s. of
eq.(5.4) is the correlation function with the gravitational descendant
$\sigma_3 (O_1) $: $<\sigma_3 (O_1) \prod_{j \in S} O_j >_g $.

\chapter{{\bf Conclusions and Discussions}}

     We have discussed the Ward identities of the Liouville gravity coupled to
the minimal CFT. We found the series of the pseudo-null fields and the
generalized equations of motion. The various Ward identities given by inserting
 these equations into correlation functions are derived. Especially for the
Ising model we give a closed set of the identities. Then the several
interesting structures similar to the matrix and the topological methods
appeared. The identities we discussed  have the similar structures to
the W and Virasoro constraints; $L_{\circ}$, $L_1 $ and $W_{-1} $ for the Ising
 model and $L_1 $ for the general case. Really the boundary-2 structure has
the same metric on the space of scaling operators as that in the other two
methods. Also the Ward identities corresponding to $W_{-1} $ constraints
require the new boundary different from the Virasoro constraints as in the
other
methods. It should be stressed that these non-linear structures are derived
from the factorization (2.15), where the intermediate states are expanded by
the normalizable Hilbert states of the Liouville and  CFT. Since the states are
 defined by using the path integral just like the Hartle-Hawking states, it
might be
expected that we also need not abandon the superposition principle in higher
dimensional quantum gravity.

   The differences from the other methods are related to the problem of
the gravitational descendants in the Liouville gravity. Our Ward identities
are closed only by the dressed operators corresponding to the Kac table of
CFT and just have the same form as the W and Virasoro constraints given by
setting the gravitational descendants outside of the Kac table to zero. The
pseudo-null fields have indeed the similar properties to the
gravitational descendants outside of the Kac table. These properties are ruled
by the Liouville charges  and the fusion rule of CFT.  Therefore the
gravitational descendants should have the same Liouville charge
and matter field as that listed in table (3.12).

   Lian and Zuckerman\NPrefmark{\lz} found a series of BRST invariant states
with these
properties   as  a candidate of gravitational descendants. For example one
can easily construct the states with the same Liouville charge and matter
field as that of $N_{1,2} $ and $N_{2,1} $
$$
\eqalign{
       R_{1,2} & = \biggl\{
            b_{-2}c_1 +{m+1 \over m}(L_{-1}^{\phi} -L_{-1}^m )\biggr\}
             \cdot {\rm e}^{\beta_{1,2}\phi} \Phi_{1,2}    ~,          \cr
       R_{2,1} & = \biggl\{
            b_{-2}c_1 +{m \over m+1}(L_{-1}^{\phi} -L_{-1}^m )\biggr\}
             \cdot {\rm e}^{\beta_{2,1}\phi} \Phi_{2,1}    ~.          \cr
         }   \eqno\eq
$$
The BRST invariance of these states are proved by using the null states of the
Liouville and the matter sectors.
Note that these states have  zero ghost number. If the measure of moduli space
is taken into account, the correlation function with these fields
vanishes by the ghost number conservation. This situation might have something
to do with that the Ward identities of the Liouville gravity have the form
mentioned above. If one wants the non-vanishing correlation functions, it is
necessary to change the measure of moduli.

  The author would like to thank T. Yoneya for careful reading of the \break
manuscript. This work is supported in part by Soryuushi Shogakukai.

\endpage

\refout

\bye